\documentstyle[12pt]{article}

\newcommand{\mybibitem}[1]{\bibitem{#1}}

\newcommand{\be}[1]{\begin{eqnarray} \label{#1} }
\newcommand{\eq}{\end{eqnarray}}

\def\a{\alpha}

\def\d{\delta}
\def\f{\phi}	
\def\vf{\varphi}
\def\bff{{\bf \Phi}}

\def\j{\psi}
\def\l{\lambda}
\def\m{\mu}

\def\p{\pi}			
\def\q{\theta}			

\def\x{\xi}

\def\F{\Phi}

\def\L{\Lambda}

\def\co{\cal{O}}

\def\half{\frac{1}{2}}
\def\pr{^{\prime}}			
\def\tr{\hbox{Tr}}			

\newcommand{\NPB}[1]{Nucl.\ Phys.\ {\bf B#1}}
\newcommand{\PLB}[1]{Phys.\ Lett.\ {\bf B#1}}

\newcommand{\PRD}[1]{Phys.\ Rev.\ {\bf D#1}}

\makeatletter
\@addtoreset{equation}{section}
\@addtoreset{equation}{subsection}
\def\theequation{\ifnum\value{section}=0 \arabic{equation}\ignorespaces
\else \ifnum\value{subsection}=0 \thesection.\arabic{equation}\ignorespaces
\else \thesection.\arabic{subsection}.\arabic{equation}\ignorespaces
                       \fi
                 \fi}
\@addtoreset{table}{section}
\@addtoreset{table}{subsection}
\def\thetable{\ifnum\value{section}=0 \arabic{table}\ignorespaces
\else \ifnum\value{subsection}=0 \thesection.\arabic{table}\ignorespaces
\else \thesection.\arabic{subsection}.\arabic{table}\ignorespaces
                    \fi
              \fi}
\makeatother

\newcommand{\pzero}{\left\{D^2,\bar{D}^2\right\}}
\newcommand{\phalf}{D^{\a}\bar{D}^2D_{\a}}
\newcommand{\xm}{d^4x}
\newcommand{\chm}{d^2\q}
\newcommand{\ftm}{d^4\q}

\begin{document}

\begin{titlepage}

\begin{flushright}
ITP-SB-96-24 \\
BRX-TH-395\\
USITP-96-07 \\
hep-th/xxyyzzz \\
\end{flushright}

\begin{center}

\vskip3em
{\large\bf Effective K\"ahler Potentials}

\vskip3em
{\normalsize M.T.\ Grisaru\footnote{E-mail:grisaru@binah.cc.brandeis.edu}\\
\vskip .5em{\it
Physics Department\\ Brandeis University \\ Waltham, MA 02254, USA}\\
\vskip2em

M.\ Ro\v cek\footnote{E-mail:rocek@insti.physics.sunysb.edu}
and R.\ von
Unge\footnote{E-mail:unge@insti.physics.sunysb.edu}${}^,$\footnote{On leave from
Department of Physics, Stockholm University, Sweden}\\
\vskip .5em
{\it Institute for Theoretical Physics\\ State University of New York\\
 Stony Brook, NY 11794-3840, USA\\}
}
\end{center}

\vfill

\begin{abstract}

\noindent
We compute the $1$-loop effective K\"ahler potential in the most general
renormalizable $N=1$ $d=4$ supersymmetric quantum field theory.
\end{abstract}

\vfill
\end{titlepage}
\section{Introduction}\label{intro}
The effective action of a theory incorporates all its quantum corrections.
It is,
however, nonlocal and difficult to calculate even in exactly solvable models.
Consequently, various approximations are used to evaluate parts of the effective
action. In particular, one frequently uses a loop expansion and a momentum
expansion. The leading term in the momentum expansion is the effective potential
\cite{cw}, and it determines the vacuum structure of the theory.  The next
term is of order $p^2$, and, together with the effective potential, determines
the masses of the states in the theory.

In $N=1$ supersymmetric theories, there is a nonrenormalization theorem that
protects the superpotential from corrections; the calculation that is directly
analogous to calculations of the effective potential is the calculation of the
effective K\"ahler potential \cite{buchart,buchbook,BMM}. When supersymmetry
is unbroken, this quantity determines both the effective potential as well
the normalization of the $p^2$ term. In this paper, we compute the
$1$-loop effective K\"ahler potential in the most general renormalizable $N=1$
$d=4$ supersymmetric quantum field theory. Our methods are considerably simpler
than earlier analyses \cite{buchart,buchbook}, and our results are more general
and complete; they take a particularly nice form in supersymmetric Landau gauge.
We find this encouraging, since it has been argued that, at least to
one-loop, the
Landau gauge effective action is equivalent to the DeWitt-Vilkovisky gauge
independent effective action \cite{dw,vi,re}.

The paper is organized as follows: After describing the most general
renormalizable $N=1$ $d=4$ supersymmetric theory, we compute the 1-loop
effective K\"ahler potential using standard super-Feynman rules in super-Landau
gauge. We then use functional methods to generalize our result to arbitrary
values of the gauge parameter.  Finally, we study some $N=2$ examples: the
contribution of a massive hypermultiplet to the vector multiplet low energy
effective action, and the effective K\"ahler potential for the hypermultiplet
itself.

\section{Setup}
We start from the most general renormalizable $N=1$ $d=4$ supersymmetric action
(we use the conventions of \cite{theBook}):
\be{N1action}
  S &=& \int \xm\chm \frac1{8g^2} W^\a_AW_\a^A
+\int \xm\ftm \bar\F e^V\F \nonumber\\
  & & +\left(\int \xm \chm (\half\F^i m_{ik}\F^k +
   \frac{1}{3!}\lambda_{ijk}\F^i\F^j\F^k) + h.c.\right)\ .
\eq
In this expression the chiral superfields $\F$ are in some product
representation of a big gauge group consisting of all the relevant gauge
groups in the problem. For example, in the $N=2$ case with a matter
hypermultiplet $Q,\tilde Q$ in a representation $R$ of the gauge group and the
chiral component superfield $\f$ of the vector multiplet,
$\F$ would be in a direct sum of the adjoint representation with
$R$ and the conjugate representation $\tilde{R}$:
\be{N=2fi}
 \F=\left(\begin{array}{c}
   \f_{A} \\ Q_{i}\\ \tilde{Q}^{k}\end{array}\right)\ .
\eq
The mass matrix $m$ as well as the couplings $\lambda$ are gauge invariant and
symmetric. Again, in the $N=2$ case, $m$ would look like
\be{N=2mass}
 m=\left(\begin{array}{ccc}
          0   & 0   & 0\\
          0   & 0   & m_{H}\d_{ik}\\
          0   & m_{H}\d_{ki} & 0
         \end{array}\right)\ ,
\eq
where $m_{H}$ is the mass of the hypermultiplet and $\l$ would be such that
$\f^{A}\l_{Aik}=\f^{A}\left(T_{A}^{R}\right)_{ik}$
and $\bar{\f}^{A}\l_{Aik}=\bar{\f}^{A}\left(T_{A}^{\tilde{R}}\right)_{ik}$

\section{Feynman graph calculation}
We now give an explicit Feynman graph calculation of the 1-loop
effective K\"ahler potential. As we shall see, this is simplest in
supersymmetric Landau gauge.

To calculate the one loop K\"ahler potential, we
expand the action to quadratic order around some constant
($x,\theta$-independent) background value $\bff$ of the scalar fields,
{\it i.e.}, $\F=\bff+\vf$. Propagators are defined from kinetic terms that are
independent of the background value $\bff$ of
$\F$, and vertices are defined from the remaining mass and $\bff$
dependent terms. It is convenient to define a $\bff$ dependent $\vf\vf$
mass $\m$
that contains all the $m$ and $\l$ dependence, as well as $\bff$ dependent
$\vf V$ and $VV$ masses $X$ and $M$:
\be{notation}
 \m_{ik} &=& m_{ik}+\l_{ijk}\bff^{j}\nonumber\\  & & \nonumber\\
 X^{A}_{i} &=& 2g\left(T_{A}\right)_{i}^{k}\bff_{k}\nonumber\\  & & \nonumber\\
 \bar{X}_{A}^{i} &=& 2g \bar{\bff}^{k}\left(T_{A}\right)_{k}^{i}\nonumber\\
  & & \nonumber\\
 M_{AB} &=& \half\left(\bar{X}_{A}^{i}X_{Bi}+\bar{X}_{B}^{i}X_{Ai}\right)\ ,
\eq
where we have rescaled $V$ by a factor $2g$.
Adding a supersymmetric gauge fixing term $\x^{-1}D^2V\bar{D}^2V$, we can write
the general gauge action as
\be{ggaction}
  \half\int \xm \ftm \left( V^A\left(\phalf-
   \frac{1}{\x} \pzero\right)V_{A} + 2\bar{\vf}^i\vf_i +
   2V^A \bar{X}^i_A\vf_i
       \right. \nonumber\\ \left.
   + 2 \bar{\vf}^{i} X_{i}^{A}V_A + V^{A}M_{AB}V^{B} \right) +
   \half\left( \int \xm \chm \vf^i\m_{ik}\vf^{k} + h.c.\right).
\eq
In principle there are also ghosts in the action, but since they do not
couple to the scalar fields, to one-loop we need not consider those terms. The
effective K\"ahler potential is most conveniently calculated in supersymmetric
Landau gauge $\x=0$. In this gauge the $VV$ propagator becomes $-\phalf/\Box^2$,
which implies that by $D$-algebra at one-loop there can be no mixed
contributions, {\it i.e.}, loops containing both $VV$ and $\vf\bar{\vf}$
propagators. This leaves only two basic loops that have to be calculated. One is
the sum over diagrams with $n$ external $\m$ vertices and $n$ external
$\bar{\m}$ vertices and $2n$ scalar field propagators $-1/\Box$. After Fourier
transforming and summing up all the diagrams we get
\be{hyperloop}
 \int \frac{d^4kd^4\q}{\left(2\p\right)^4}\frac{1}{2k^2}
   \tr\ln\left(1+\frac{\bar{\m}\m}{k^2}\right)=
  \frac{1}{2\left(4\p\right)^2}\tr\int d^4\q dk^2
    \ln\left(1+\frac{\bar{\m}\m}{k^2}\right)\ .
\eq
The other is the sum over diagrams with $n$ external $M$ vertices
and n vector propagators $-\phalf/\Box$ which, after Fourier
transforming and doing the sum, gives
\be{vectorloop}
 -\int \frac{d^4kd^4\q}{\left(2\p\right)^4}\frac{1}{k^2}
   \tr\ln\left(1+\frac{M}{k^2}\right)=
  -\frac{1}{\left(4\p\right)^2}\tr\int d^4\q dk^2
    \ln\left(1+\frac{M}{k^2}\right)\ .
\eq
Evaluating the momentum integral with an ultraviolet cut-off $\Lambda$, we find
the regularized 1-loop effective K\"ahler potential for the most general
renormalizable four-dimensional theory:
\be{result}
K_{eff}=-\frac{1}{2\left(4\p\right)^2}\tr\left[\bar\m \m\ln(\frac{\bar\m\m}
{\hbox{e}\Lambda^2})-2M\ln(\frac M{\hbox{e}\Lambda^2})\right]\ .
\eq
Note that the factor of e $=\exp(1)$ can be removed by a finite renormalization.

\section{General gauge}
The effective action, being an off-shell quantity, is not gauge independent. For
completeness, we now compute the 1-loop effective K\"ahler potential for
arbitrary gauge fixing parameter $\xi$ using functional methods. To
functionally integrate over the chiral fields, it is convenient to solve the
chirality constraint on
$\vf$ and $\bar{\vf}$ by introducing unconstrained fields $\j$ and $\bar{\j}$
such that $\vf=\bar{D}^2\j$ and $\bar{\vf}=D^2\bar{\j}$. In principle this
introduces a new gauge invariance into the action, but in the
absence of background gauge fields $V$, the ghosts associated with this gauge
fixing decouple (actually, covariant functional methods can also be used in a
$V$ background, but they are not needed here \cite{theBook,gz}). The
unconstrained functional integration over $V,\j$ and $\bar{\j}$ gives a
contribution to the effective action:
\be{effact}
 -\half\int \xm\ftm\ftm^{\pr}
   \d(\q^{\pr}-\q)\tr\, \ln{\co} (x,\q)\d(\q-\q^{\pr})\ ,
\eq
where the $-1/2$ comes from the fact that we are computing $\det^{-\half}$
and $\co$ is the kinetic operator
\be{matrix}
\co=\left( \begin{array}{ccc}
       \phalf - \frac1\x\pzero + M &
       \bar{X}\bar{D}^2 & XD^2 \\ & & \\
       \bar{X}\bar{D}^2 & \m\bar{D}^2 & \Box\\ & & \\
       XD^2 & \Box & \bar{\m}D^2
       \end{array}
\right)\ .
\eq
The logarithm can be split into two pieces using the formula for the
determinant of a block matrix
\be{block}
\det\left(\begin{array}{cc}
              A & B\\
              C & E\end{array}\right)=
       \det\left(E\right)\det\left(A-B\,E^{-1}\,C\right)\ .
\eq
To do this we need the formula for the inverse of $E$ which in our
case is
\be{Einv}
E^{-1}=
\left(\begin{array}{cc}
        \m\bar{D}^2 & \Box\\
        \Box & \bar{\m}D^2
      \end{array}\right)^{-1}
    = \frac{1}{\Box}\left(\begin{array}{cc}
 -\bar{\m}\frac{D^2}{\Box-\m\bar{\m}} &
 1+\bar{\m}\m\frac{D^2\bar{D}^2}{\Box(\Box-\bar{\m}\m)}\\ & \\
 1+\m\bar{\m}\frac{\bar{D}^2D^2}{\Box(\Box-\m\bar{\m})}&
 -\m\frac{\bar{D}^2}{\Box-\bar{\m}\m}\end{array}\right)\ .
\eq
Using this, we can write
\be{split}
\tr\ln\left(\co\right) &=& \tr\ln\left(E\right)\nonumber\\ &&\nonumber\\ &&+
~\tr\ln\left(\phalf-\frac{1}{\x}\pzero
 +M_{AB}\right.\nonumber\\ &&\left.\right.\nonumber\\ &&\left.\qquad
  -\frac{\bar{D}^2D^2}{\Box}\bar{X}^{i}_{A}X_{Bi}-
 \frac{D^2\bar{D}^2}{\Box}
 X^{i}_{A}\bar{X}_{Bi}
 \right. \nonumber\\ &&\left.\right.\nonumber\\ &&\left.\qquad
 -\frac{\bar{D}^2D^2}{\Box}
 \bar{X}^{i}_{A}\left[\bar{\m}\m\frac{1}{\Box-\bar{\m}\m}\right]_{ik}X^{k}_{B}
 \right. \nonumber\\ &&\left.\right.\nonumber\\ &&\left.\qquad
 -\frac{D^2\bar{D}^2}{\Box}X^{i}_{A}
 \left[\m\bar{\m}\frac{1}{\Box-\m\bar{\m}}\right]_{ik}\bar{X}^{k}_{B}
 \right. \nonumber\\ &&\left.\right.\nonumber\\ &&\left.\qquad
 +\bar{D}^{2}\bar{X}^{i}_{A}\left[\bar{\m}\frac{1}{\Box-\m\bar{\m}}\right]_{ik}
 \bar{X}^{k}_{B}+D^2X^{i}_{A}\left[\m\frac{1}{\Box-\bar{\m}\m}\right]_{ik}
  X^{k}_{B}
\right)\ .\nonumber\\ &&
\eq
Since we are only interested in the $\bff$ dependence, we can
factor out the $\bff$ independent $V$-propagator piece and subsequently
drop it. Doing this and using the symmetry of $\m$ and $\bar{\m}$ we
have
\be{before_alg}
\tr\ln\left(\co\right) &=& \tr\ln\left(E\right)\nonumber\\
 &&+~\tr\ln\left(1+\frac{\phalf}{\Box^{2}}M_{AB}+
 \x\frac{\bar{D}^2D^2}{\Box^{2}}\left(S_{AB}+T_{AB}\right)
 \right. \nonumber\\ &&~~~~~~~~\left.
 +~\x\frac{D^2\bar{D}^2}{\Box^2}\left(S_{BA}+T_{BA}\right)-
 \x\frac{D^2}{\Box}R_{AB}-\x\frac{\bar{D}^2}
{\Box}\bar{R}_{AB}\right)\ ,\nonumber\\
\eq
where
\be{matdefs}
S_{AB} &=& \half\left(\bar{X}^{i}_{A}X_{Bi}-\bar{X}^{i}_{B}X_{iA}\right)
 \nonumber\ ,\\ & & \nonumber \\
T_{AB} &=& \bar{X}^{i}_{A}\left[\bar{\m}\m\frac{1}{\Box-\bar{\m}\m}\right]_{ik}
  X^{k}_{B}\nonumber \ ,\\ & & \nonumber \\
R_{AB} &=& \half X^{i}_{(A}\left[\m\frac{1}{\Box-\bar{\m}\m}\right]_{ik}
  X^{k}_{B)}\ .\nonumber\\
\eq
Because the operator $\phalf$ annihilates anything
proportional to $D^2$ or $\bar{D}^2$, this second trace actually splits up
into a sum of two terms, and we can thus write (\ref{split}) as a sum of three
terms
\be{threeterms}
 \tr\ln\co &=& \tr\ln\left(E\right)
 +\tr\ln\left\{1+\frac{\phalf}{\Box^{2}}M_{AB}\right\}
 \nonumber\\
 &&+\tr\ln\left\{1+\x\left(\frac{\bar{D}^2D^2}{\Box^2}\left(S_{AB}+
 T_{AB}\right)
 \right.\right. \nonumber\\ &&\left.\left.
 +\frac{D^2\bar{D}^2}{\Box^2}\left(S_{BA}+T_{BA}\right)-
 \frac{D^2}{\Box}R_{AB}-\frac{\bar{D}^2}{\Box}\bar{R}_{AB}\right)\right\}\ .
\eq

We now want to do the $D$-algebra in the expression (\ref{effact}).
To this end we insert $1$ in the form $\left(\pzero-\phalf\right)/\Box$ in
front of $\tr\ln\co$ (see Eq.~(\ref{firstbefore}), below).  Using the properties
of the projection operators we can convert all the spinor derivatives inside the
logarithms into boxes. Let us analyze each term separately.

We extract an irrelevant factor $\propto\tr\ln(\Box)$ from the first term (and
drop it in Eq.~\ref{rewriting} and below):
\be{firstfudge}
\tr\ln(E)= \tr\ln\left(\begin{array}{cc} 0 & \Box\\ \Box & 0
\end{array}\right)+
\tr\ln\left\{\left(\begin{array}{cc} 1 & 0\\ 0 & 1
\end{array}\right)
 +\left(\begin{array}{cc} 0 & \frac{\bar{\m}D^2}{\Box}\\
                  \frac{\m\bar{D}^2}{\Box} & 0 \end{array}\right)\right\}\ .
\eq
Because of the trace, only even powers survive in the expansion of the
log and we can rewrite (\ref{firstfudge}) as
\be{rewriting}
 \half\tr\ln\left\{\left(\begin{array}{cc} 1 & 0\\ 0 & 1
                                  \end{array}\right)
 -\left(\begin{array}{cc} \frac{\bar{\m}\m D^2\bar{D}^2}{\Box^2} & 0\\
              0 & \frac{\m\bar{\m}\bar{D}^2D^2}{\Box^2}\end{array}
   \right)\right\}\ .
\eq
Performing part of the trace and inserting $1$ in the form given
above, the final expression is:
\be{firstbefore}
 -\half\int \xm\ftm\ftm^{\pr}
   \d\left(\q^{\pr}-\q\right)\tr\ln\left(E\left(x,\q\right)\right)
   \d\left(\q-\q^{\pr}\right)=\\ \nonumber
 -\frac{1}{4} \int \xm\ftm\ftm^{\pr}
   \d\left(\q^{\pr}-\q\right)\frac{\pzero-\phalf}{\Box}\times\nonumber\\
   \times\tr\ln\left(1-\frac{D^2\bar{D}^2}{\Box^2}\bar{\m}\m
     -\frac{\bar{D}^2D^2}{\Box^2}\m\bar{\m}\right)
   \d\left(\q-\q^{\pr}\right)\ .
\eq
After doing the $D$-algebra and using the properties of the spinor
derivatives and the delta functions, and taking the
Fourier transform, this reduces to
\be{firstafter}
 \frac{1}{2\left(4\p\right)^2} \int\ftm dk^2
  \tr\ln\left(1+\frac{\bar{\m}\m}{k^2}\right)\ ,
\eq
where the final trace is over the $i,k$ indices of the $\bar{\m}\m$
matrix and this agrees with the scalar multiplet contribution
calculated in super-Landau gauge (\ref{hyperloop}).

The second term, treated in the same way, becomes
\be{secondafter}
 -\frac{1}{\left(4\p\right)^2}\int \ftm dk^2
  \tr\ln\left(1+\frac{M}{k^2}\right)\ ,
\eq
where the trace is over the adjoint indices of $M_{AB}$; this
agrees with our previous (super-Landau gauge) result (\ref{vectorloop}).

The third term contains all the $\x$ dependence and in Landau gauge it
vanishes. The $D$-algebra in this
case is not so straightforward as in the previous cases. To perform it
we have to use the identity
\be{useful}
 \lefteqn{1+XD^2\bar{D}^2+Y\bar{D}^2D^2+ZD^2+\bar{Z}\bar{D}^2 =}&
 \nonumber\\&\nonumber\\
 &\left(1+\bar{N}\bar{D}^2\right)\left(1+UD^2\bar{D}^2+V\bar{D}^2D^2\right)
  \left(1+ND^2\right)\ ,
\eq
where $X,Y,Z,\bar{Z}$ are some arbitrary matrices and
\be{newback}
 N &=& \left(1+X\Box\right)^{-1}Z\nonumber\\
 \bar{N} &=& \bar{Z}\left(1+X\Box\right)^{-1}\nonumber\\
 U &=& X\\
 V &=& Y-\bar{Z}\left(1+X\Box\right)^{-1}Z\ .\nonumber
\eq
The $D$-algebra implies that a function of $D^2$ or $\bar{D}^2$ alone vanishes,
and we finally get all of the $\x$ dependence as a sum of two terms.
\be{finally}
 \lefteqn{\frac{1}{2\left(4\p\right)^2}\int \ftm dk^2 \left\{
  \tr\ln\left(1-\x\frac{1}{k^2}\left(S+T\right)_{AB}\right)
  \right.}& \nonumber\\ &\left.
   +\tr\ln\left(1-\x\frac{1}{k^2}\left(S+T\right)_{AB}+
  \x^2\bar{R}_{AC}\left[\frac{1}{k^2-\x\left(S+T\right)
  }\right]_{CD}R_{DB}
  \right)\right\}\ .
\eq
We have verified that when $\bff$ is restricted to fields which, after
gauge symmetry breaking, are massless and hence on shell at zero
momentum, the $\x$ dependent terms vanish. Unfortunately, it seems
impossible to actually evaluate the momentum integral and find an explicit
expression for the $\x$ dependent term in the most general case.

\section{Examples}
We now consider some examples; we focus on the
$N=2$ case, where our field $\F$ contains an adjoint scalar $\f$ (which we
can regard as a matrix) and a number of hypermultiplets $Q,\tilde{Q}$. First
consider the case with one hypermultiplet with mass $m$ and let us look at the
$\f$ dependence of the effective K\"{a}hler potential. The
hypermultiplet contribution (\ref{firstafter}) gives two equal
factors summing up to
\be{massivehyper}
 \frac{1}{\left(4\p\right)^2}\int dk^2 \tr\ln\left(1+
  \frac{\left(\bar{m}+\bar{\f}\right)\left(m+\f\right)}{k^2}\right)
\eq
where the trace is over the representation of the hypermultiplet.
The vector multiplet contribution from (\ref{secondafter}) is
\be{massivevector}
 -\frac{1}{\left(4\p\right)^2}\int dk^2 \tr\ln\left(1+
  \frac{\f\bar{\f}+\bar{\f}\f}{2k^2}\right)
\eq
where the trace is over the adjoint representation.
Since we are only considering the dependence on $\f$, the matrices $T$ and
$R$ are zero. However $S$ (see \ref{matdefs}) is not zero and this gives us a
$\x$ dependent K\"{a}hler potential where the $\x$ dependent piece is
\be{massivexi}
 \frac{1}{\left(4\p\right)^2}\int dk^2 \tr\ln\left(1+
  \x\frac{\f\bar{\f}-\bar{\f}\f}{2k^2}\right)
\eq
We can perform the integral and the trace in each case. The vector
multiplet contribution is the same as in \cite{BMM} but the
hypermultiplet contribution changes to
\be{massivetotal}
 K^{fund}_{Q} =
 -\frac{1}{\left(8\p\right)^2}\left(s\ln\frac{\left(\bar{\f}^2-\bar{m}^2\right)
 \left(\f^2-m^2\right)}{16\L^4}+t\ln\frac{s+t}{s-t}\right)\ ,
\eq
where $s=\bar{m}m+\bar{\f}\cdot\f$ and $t=\sqrt{(\bar{m}\f+m\bar{\f})\cdot
(\bar{m}\f+m\bar{\f})+(\bar{\f}\cdot\f)^2-\bar{\f}^2\f^2}$.
Restricting $\f$ to the massless fields, {\it i.e.,} $[\f,\bar\f]=0$,
this agrees with the expression one gets by expanding the exact solution of
Seiberg and Witten \cite{sw}.\footnote{We thank Gordon Chalmers for evaluating
the asymptotic expansions of the relevant elliptic integrals.}

We find the dependence on the gauge fixing parameter by evaluating the integral
and the trace in (\ref{massivexi}). The eigenvalues of the matrix
$[\f ,\bar{\f}]$ are $\left\{ 0,~\pm
\sqrt{\left(\bar{\f}\cdot\f\right)^2-\bar{\f}^2\f^2}\right\}$,  so when doing
the $k^2$ integral, we are left with an {\em imaginary}  contribution to the
effective action $i\p\x\sqrt{\left(\bar{\f}\cdot\f\right)^2-\bar{\f}^2\f^2}$.
Such a term in the effective action may appear disturbing, but it is typical of
unphysical contributions from longitudinal states that are projected out in
Landau gauge.\footnote{We thank George Sterman for explaining this to us.}
Indeed, it does not contribute to physical quantities, and, in particular, it
vanishes for the massless fields of the theory (when $\f$ and $\bar\f$ commute).
It also does not arise in the gauge-independent effective action of DeWitt and
Vilkovisky
\cite{dw,vi,re}.

Next we consider $N_{f}$ hypermultiplets in an arbitrary
representation and with $m=0$, and study the dependence of the
K\"{a}hler potential on the hypermultiplets. For simplicity we also
work in super Landau gauge ($\x$=0). The hypermultiplet
contribution again is a sum of two equal factors giving
\be{nfhyper}
 \frac{1}{\left(4\p\right)^2}\int dk^2 \tr\ln\left(\d_{AB}+
  \frac{\sum_{a=1}^{N_{f}} \left(\bar{Q}^{a}T_{A}T_{B}Q^{a}+
   \tilde{Q}^{a}T_{B}T_{A}\bar{\tilde{Q}^{a}}\right)}{k^2}\right)\ ,
\eq
and the vector multiplet contribution is
\be{nfvector}
 -\frac{1}{\left(4\p\right)^2}\int dk^2 \tr\ln\left(\d_{AB}+
  \frac{\sum_{a=1}^{N_{f}} \left(\bar{Q}^{a}
      \left\{T_{A},T_{B}\right\}Q^{a}+\tilde{Q}^{a}
      \left\{T_{A},T_{B}\right\}\bar{\tilde{Q}^{a}}\right)}{2k^2}\right)\ .
\eq
Note that $T_AT_B=\half(\{T_A,T_B\}+[T_A,T_B])=\half(\{T_A,T_B\}+f_{AB}^CT_C)$
implies that the two contributions cancel exactly whenever
\be{vacuum}
\sum_{a=1}^{N_{f}} \left(\bar{Q}^{a}T_{A}Q^{a}-
\tilde{Q}^{a}T_{A}\bar{\tilde{Q}^{a}}\right)=0\ ;
\eq
This condition is satisfied by hypermultiplets that remain massless.
Thus, massless hypermultiplets receive no corrections to their K\"ahler
potential, as stated in \cite{sw}. However, the massive multiplets are expected
to receive corrections, and, in particular, we expect to find terms that mix
the hypermultiplets and the vector multiplet chiral field $\f$.  These are
presumably K\"ahler residues of higher dimension terms analogous to those found
for the vector multiplet alone in \cite{BMM}; however, since a suitable
off-shell formulation of the hypermultiplet is not known, we do not even know
how to write such terms down.

While we were preparing this manuscript, a preprint \cite{West} appeared on the
hep-th archive; the work has some overlap both with this paper and with
\cite{BMM}. The authors do not discuss their gauge choice, but appear to work in
supersymmetric Feynman gauge without considering the imaginary term  that we
found away from Landau gauge.

\end{document}